\documentclass[aps,prl,floatfix,showpacs,twocolumn,longbibliography,superscriptaddress]{revtex4-1}
\usepackage{ulem}
\usepackage{dcolumn}
\usepackage{bm}
\usepackage{color}
\usepackage{amssymb}
\usepackage{amsmath}
\usepackage{graphicx}
\usepackage{amsfonts}
\usepackage{slashed}
\usepackage{pstricks}
\usepackage{float}
\usepackage{multirow}
\usepackage{array}
\allowdisplaybreaks
\usepackage{dcolumn}
\usepackage{epsf}
\usepackage{float}
\usepackage[nice]{nicefrac}

\newcommand{\bra}{\langle}
\newcommand{\ket}{\rangle}

\begin{document}
\title{Distilling the essential elements of nuclear binding via neural-network quantum states}

\author{Alex Gnech}
\affiliation{European Center for Theoretical Studies in Nuclear Physics and Related Areas (ECT*)
and Fondazione Bruno Kessler Strada delle Tabarelle 286, I-38123 Villazzano (TN), Italy}
\affiliation{INFN-TIFPA Trento Institute of Fundamental Physics and Applications, 38123 Trento, Italy}

\author{Bryce Fore}
\affiliation{Physics Division, Argonne National Laboratory, Argonne, IL 60439, USA}

\author{Alessandro Lovato}
\affiliation{INFN-TIFPA Trento Institute of Fundamental Physics and Applications, 38123 Trento, Italy}
\affiliation{Physics Division, Argonne National Laboratory, Argonne, IL 60439, USA}
\affiliation{Computational Science Division, Argonne National Laboratory, Argonne, IL 60439, USA}

\date{\today}
\begin{abstract}
In pursuing the essential elements of nuclear binding, we compute ground-state properties of atomic nuclei with up to $A=20$ nucleons, using as input a leading order pionless effective field theory Hamiltonian. A variational Monte Carlo method based on a new, highly-expressive, neural-network quantum state ansatz is employed to solve the many-body Schr\"odinger equation in a systematically improvable fashion. In addition to binding energies and charge radii, we accurately evaluate the magnetic moments of these nuclei, as they reveal the self-emergence of the shell structure, which is not {\it a priori} encoded in the neural-network ansatz. To this aim, we introduce a novel computational protocol based on adding an external magnetic field to the nuclear Hamiltonian, which allows the neural network to learn the preferred polarization of the nucleus within the given magnetic field.
\end{abstract} 
%
%
\maketitle

\paragraph{Introduction.---} 
Understanding emerging nuclear properties from the interactions among protons and neutrons is a longstanding challenge of nuclear theory~\cite{Hergert:2020bxy}. To address this complex problem, accurate nuclear Hamiltonians have been constructed either within phenomenological approaches~\cite{Wiringa:1994wb, Machleidt:2000ge} or by leveraging the broken chiral symmetry of QCD~\cite{Entem:2003ft, Epelbaum:2004fk, Gezerlis:2013ipa, Lynn:2015jua, Epelbaum:2014sza, Piarulli:2014bda, Entem:2017gor, Piarulli:2017dwd}. Sophisticated potentials are spin-isospin dependent, typically entail high-momentum components, and involve several fitting parameters, usually determined on light nuclear systems.

Quantum many-body methods solve the Schr\"odinger equation associated with these Hamiltonians with controlled approximations, leading to remarkable success in computing nuclei across the nuclear chart~\cite{Barrett:2013nh,Hagen:2013nca,Hergert:2015awm,Carbone:2013eqa,Epelbaum:2011md,Carlson:2014vla,Morris:2017vxi,Hu:2021trw}. However, important questions remain unanswered. For instance, no existing Hamiltonian can simultaneously reproduce with high precision the properties of light nuclear systems, the charge radii of medium-mass nuclei, and the neutron-matter equation of state~\cite{Binder:2013xaa,Lonardoni:2017egu, Sammarruca:2021bpn, Nosyk:2021pxb, Lovato:2022apd}. 

The quest for distilling the ``essential elements of nuclear binding'' has arisen from these challenges, seeking the simplest nuclear Hamiltonian yielding binding energies and charge radii across the nuclear chart with few percent errors~\cite{Lu:2018bat}. Arguments based on the unitary limit, large $N_c$, and SU(4) symmetry~\cite{Kievsky:2018xsl, Lu:2018bat, Lee:2020esp, Kievsky:2021ghz} suggest that such a Hamiltonian could be derived within pionless effective field theory ($\slashed{\pi}$EFT)~\cite{Bedaque:2002mn}. In Ref.~\cite{Lu:2018bat} it was proven that a SU(4)-symmetric Hamiltonian, with a non-local three-body force, can accurately model ground-state properties of several nuclei up to $A=40$ nucleons and neutron matter up to saturation density. Similarly, the leading-order (LO) $\slashed{\pi}$EFT Hamiltonian ``o'' developed in Ref.~\cite{Schiavilla:2021dun} reproduces reasonably well the binding energies of various closed-shell nuclei and yields a neutron matter equation of state that is remarkably close to realistic Hamiltonians~\cite{Fore:2022ljl} at low densities. 

In this work, we carry out an in-depth assessment of the proposed essential Hamiltonians by computing ground-state properties of nuclei with up to $A=20$ nucleons. In addition to their binding energies and radii, we analyze their magnetic moments, which determine the interactions of nuclei with external magnetic fields and contribute to the hyperfine structure in the electronic spectra of the atom~\cite{Emery:2006}. Owing to the significant difference between the $g$ factors associated with the orbital and spin angular momenta of protons and neutrons, they are an ideal test-bed for nuclear models~\cite{Bohr:1998}.

Variational Monte Carlo (VMC) and Green's function Monte Carlo (GFMC) calculations of light nuclei carried out with realistic Hamiltonians demonstrated the need for two-body contributions in the electromagnetic currents~\cite{Marcucci:2008mg, Pastore:2012rp}. Owing to the exponential growth of the spin-isospin basis~\cite{Carlson:2014vla}, the GFMC is limited to $A\leq 12$ nuclei. The authors of Ref.~\cite{Martin:2023dhl} used the auxiliary-field diffusion Monte Carlo (AFDMC) to evaluate magnetic moments of nuclei up to $^{17}$F using chiral-EFT Hamiltonians and consistent currents. However, the AFDMC cannot reach much larger systems, owing to the fermion-sign problem, which is exacerbated by the use of oversimplified variational wave functions~\cite{Gandolfi:2014ewa}.

To overcome these limitations, building on the success of earlier works~\cite{Keeble:2019bkv, Adams:2020aax, Gnech:2021wfn, Yang:2022esu}, we solve the nuclear many-body Schr\"odinger equation in a systematically-improvable fashion using a VMC method based on a neural-network quantum state (NQS) ansatz. The expressivity of the hidden-nucleon NQS, initially developed to model the ground-state wave functions of $^3$H, $^3$He, $^4$He, and $^{16}$O~\cite{Lovato:2022tjh}, and dilute neutron matter~\cite{Fore:2022ljl}, is augmented through neural backflow transformations~\cite{Luo:2019}. The latter are realized by a simplified version of the permutation-equivariant message-passing neural network (MPNN), recently adopted to compute the homogeneous electron gas~\cite{Pescia:2023mcc} and cold Fermi gases~\cite{Kim:2023fwy}. To reliably evaluate the nuclear magnetic moments, we introduce a novel computational protocol that allows the NQS to learn the preferred polarization of the nucleus. 

\paragraph{Methods.---} 
We model the interactions among protons and neutrons with the LO $\slashed{\pi}$EFT Hamiltonian, ``o'', developed in  Ref.~\cite{Schiavilla:2021dun}.
The nucleon-nucleon potential reproduces the $np$ scattering lengths and effective ranges in the $S/T = 0/1$ and $1/0$ channels, and it vanishes in odd partial waves. We assume the electromagnetic component to only include the Coulomb force between finite-size protons. A repulsive three-body force is needed to stabilize the systems with more than two nucleons against the Thomas collapse. AFDMC and VMC-NQS calculations showed that the choice $R_3=1.0$ fm for the three-nucleon regulator overbinds $^{16}$O and heavier nuclei~\cite{Schiavilla:2021dun,Lovato:2022tjh}. To counter it, we opt for $R_3= 1.1$ fm, as the extended range introduces additional repulsion in heavier systems.

We introduce $X=\{x_1\dots x_A\}$ to denote the set of single-particle coordinates $x_i = \{\mathbf{r}_i, s^z_i, t^z_i\}$, which describe the spatial positions and the z-projection of the spin-isospin degrees of freedom of the $A$ nucleons. The hidden-nucleon wave function~\cite{Moreno2022,Lovato:2022tjh} reads
\begin{equation}
    \Psi_{HN}(X) \equiv \text{det}\left[ 
    \begin{matrix}
        \phi_v(X) & \phi_v(X_h)\\
        \phi_h(X) & \phi_h(X_h)
    \end{matrix}
    \right]\,,
    \label{eq:psi_hn}
\end{equation}
where, $\phi_v$ and $\phi_h$ denote the visible and hidden orbitals, while $X$ and $X_h$ are the $A$ visible and the $A_h$ hidden coordinates. Hence, the dimension of the sub-matrices $\phi_v(X)$, $\phi_v(X_h)$, $\phi_h(X)$, and $\phi_h(X_h)$ are $A \times A$, $A_h \times A$, $A \times A_h$, and $A_h \times A_h$, respectively. As a major departure from Ref.~\cite{Lovato:2022tjh}, all the above matrices are complex valued, and two separate deep neural networks with differentiable activation functions parametrize the logarithm of their moduli and phases. To respect fermion anti-symmetry, the coordinates of the hidden nucleons $X_h$ are permutation-invariant functions of the visible ones. We enforce this symmetry using a Deep-Sets architecture~\cite{Zaheer:2017,Wagstaff:2019} with {\it logsumexp} pooling.  

As in recent neutron-matter studies~\cite{Fore:2022ljl}, we improve the flexibility of the ansatz by applying equivariant backflow transformations to pre-process the single-nucleon coordinates and include correlation effects. These transformations are implemented by means of a simplified version of the MPNN employed in Refs.~\cite{Pescia:2023mcc,Kim:2023fwy} 
\begin{equation}
\mathbf{y}_i = h\Big(\mathbf{x}_i, \sum_j m(\mathbf{x}_i, \mathbf{x}_j) \Big)\, .
\end{equation}

Due to the universality of the hidden-nucleon ansatz, the single, albeit enlarged Slater determinant in Eq.\eqref{eq:psi_hn} is sufficient for modeling the ground-state wave function of both closed- and open-shell nuclei, regardless of their deformation. This characteristic represents a significant advantage compared to ``conventional'' quantum Monte Carlo methods, where multiple Slater determinants are required to model open-shell systems~\cite{Lonardoni:2018nob,Gandolfi:2020pbj}. In stark contrast with the majority of nuclear many-body methods, the shell structure of the nucleus is not directly encoded in the NQS, as the parameters of the network are randomly initialized and no pre-training on a Hartree-Fock wave function is performed. All ground-state properties self-emerge during the training of the network, which is performed by minimizing the variational energy of the system. For this purpose, we employ the stochastic reconfiguration algorithm~\cite{Sorella:2005} with regularization based on the RMSprop method~\cite{Lovato:2022tjh}. The expectation value of the Hamiltonian and other quantum-mechanical operators of interest is evaluated stochastically using the Metropolis-Hastings algorithm detailed in the Supplemental Material of Ref.~\cite{Adams:2020aax}, sampling both the spatial and spin-isospin coordinates of the $A$ nucleon

\paragraph{Results.---}
\begin{figure}[t]
    \centering
    \includegraphics[width=\linewidth]{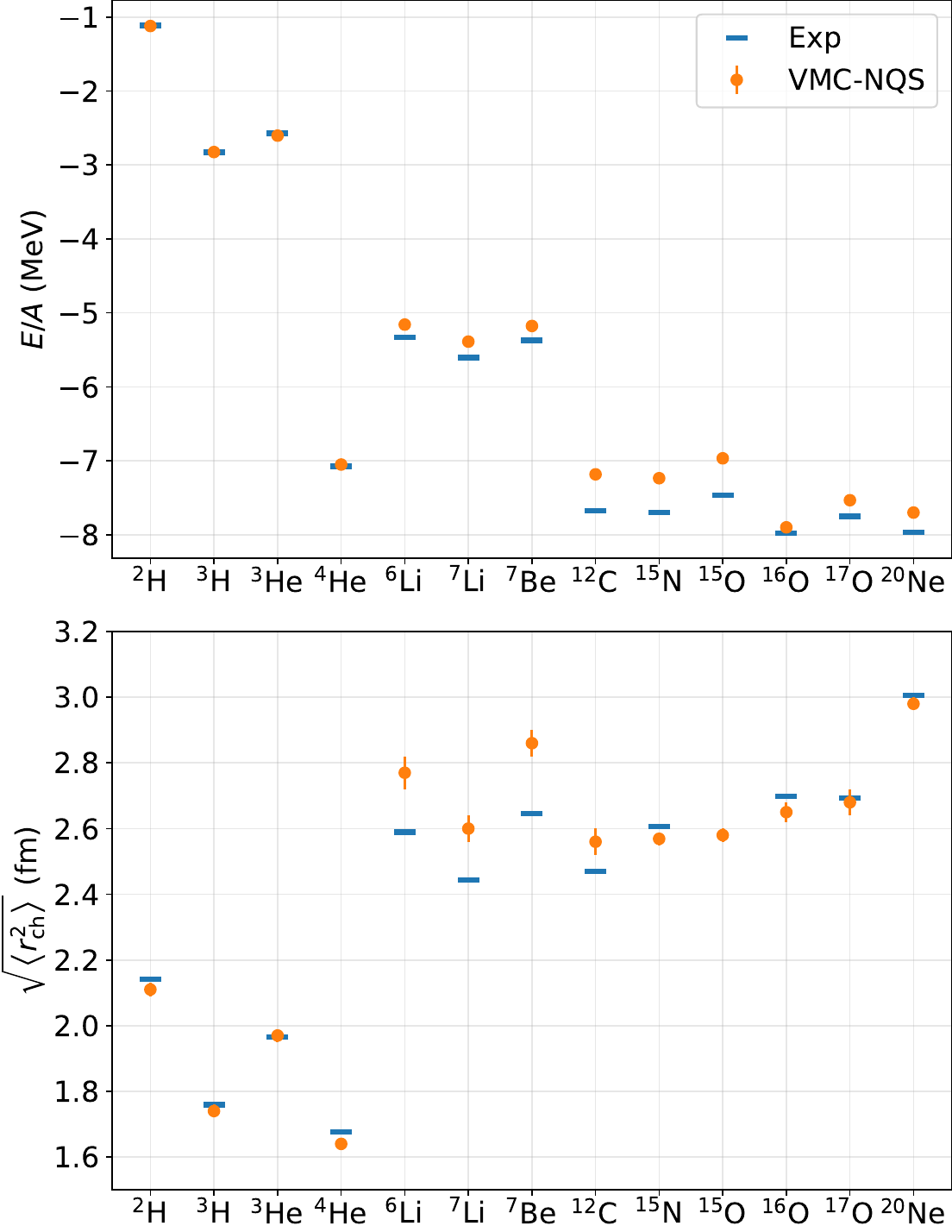}
  \caption{VMC-NQS energy per particle (upper panel) and charge radii (lower panel) of selected nuclei with up to $A=20$ nucleons as obtained from the LO EFT Hamiltonian ``o'' of Ref.~\cite{Schiavilla:2021dun} with $R_3=1.1$ fm compared with experimental data.}
  \label{fig:spectrum_radii}
\end{figure}
The upper panel of Fig.~\ref{fig:spectrum_radii} displays the ground-state energies per nucleon of selected $A\leq 20$ nuclei obtained solving the ground-state of the $\slashed{\pi}$EFT Hamiltonian with the VMC-NQS method. The agreement between the computed and experimental values is remarkably good, given the simplicity of the $\slashed{\pi}$EFT input Hamiltonian. Notably, our ground-state energies are closer to experimental values than those obtained in Ref.~\cite{Martin:2023dhl} with the AFDMC method and N$^2$LO chiral-EFT interactions. Additionally, we do not observe the increasing overbinding with the number of nucleons reported in no-core shell model calculations that use as input different N$^2$LO two- and three-body forces~\cite{Maris:2020qne}. Nevertheless, $^{12}$C and $^{15}$O are under-bound by about $0.6$ MeV per nucleon, while $^6$He, $^8$Li, $^8$B, $^9$C, and $^{17}$F are unstable against breakup into smaller clusters. This behavior suggests an excessive repulsion from the input Hamiltonian, as increasing the flexibility of the NQS by considering more hidden nucleons does not significantly improve the variational energies. 

In the lower panel of Fig.~\ref{fig:spectrum_radii}, we show the charge radii of the same nuclei, which are are derived from the estimates for the point-proton radii taking into account the finite-size of the nucleons and the Darwin-Foldy correction. Here we neglect both the one-body spin-orbit correction of Ref.~\cite{Ong:2010gf} and two-body terms in the charge operator~\cite{Lovato:2013cua}. 
 While the overall trend of experimental data is well reproduced, $^2$H, $^4$He, and $^{16}$O are slightly smaller than experiments. On the other hand, 
consistent with the ground-state energies, the radii of $^6$Li, $^7$Li, $^7$Be, and $^{12}$C are too large --- there is no available experimental data for $^{15}$O. In contrast to many-body methods relying on the harmonic-oscillator basis expansion~\cite{Caprio:2022mkg}, the radius converges quickly in our VMC-NQS calculations due to the stability of the method. Hence, the discrepancies between theoretical calculations and experimental data are likely to be ascribed to deficiencies in the input LO $\slashed{\pi}$EFT Hamiltonian. 

At LO in the electroweak current operator, the expectation value of the magnetic moment is given by
\begin{equation}
\mu = \frac{\langle \Psi_{HN} | \sum_i P_i^p L_i^z + (\mu_S+\mu_V\tau_i^z) \sigma_i^z | \Psi_{HN} \rangle}{\langle \Psi_{HN} | \Psi_{HN} \rangle }\,.
\label{eq:mag_mom}
\end{equation}
In the above equation, $ L_i^z = -i(\mathbf{r}_i\times \boldsymbol{\nabla}_i)_z$ is the projection of the orbital angular momentum on the $z$ axis, while $\mu_S=0.8798\,\mu_N$ and $\mu_V=4.0759\,\mu_N$, and $\mu_N$ is the nuclear magneton. The above definition assumes that the wave function of the nucleus has well defined total angular momentum $J$ and maximal projection on the $z$ axis: $J_z = J$. Since the LO $\slashed{\pi}$EFT Hamiltonian that we employ does not contain tensor or spin-orbit interactions, its ground state is an eigenstate of the total spin $S$ and total orbital angular momentum $L$ and it is degenerate in their projections $S_z$ and $L_z$: $|\Psi_{HN}; L,S\rangle=\sum_{L_z S_z}c^{L,S}_{L_z S_z}|L,L_z;S,S_z\rangle$. 

\begin{figure}[htb]
    \centering
    \includegraphics[width=\columnwidth]{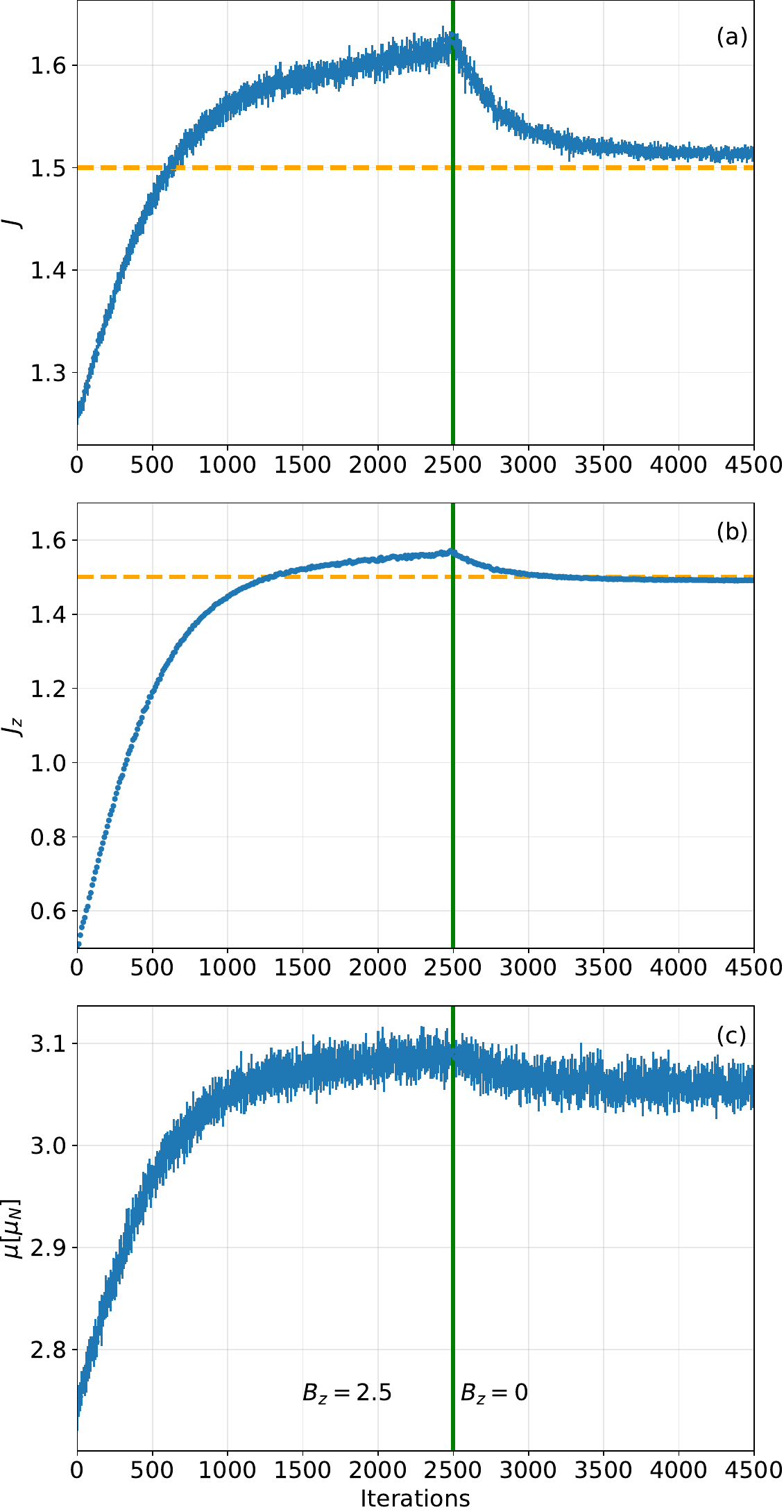}
    \caption{Total angular momentum, its projection on the $z$-axis, and the magnetic moment of $^7$Li while projecting the NQS onto the state with $J,J_z=3/2,3/2$. The magnetic field $B_z=2.5$ MeV is switched off after 2500 iterations as indicated by the vertical green line. The orange dashed lines represent the target values $J=3/2$ and, $J_z=3/2$.}
    \label{fig:B_field}
\end{figure}

The total polarization of the system $S_z$ is conveniently fixed in the Metropolis Hastings Sampling. Hence, since the ground states of $^2$H, $^3$H, $^3$He, and $^6$Li are all $L=0$ (and hence $L_z=0$), they are also eigenstates of $J$ and $J_z$. On the other hand, evaluating the magnetic moments of heavier nuclei requires projecting the NQS onto a given value of $L_z$. To this aim, we have devised a three-stage algorithm. i) A standard SR minimization of the variational energy is carried out to find a good approximation for the nuclear ground state. ii) A small external magnetic field is added to the Hamiltonian, $H \to H - B_z L_z$, and another energy minimization is carried out to resolve the degeneracy favoring larger values of $L_z$. iii) The magnetic field is switched off, and additional SR steps are performed to remove the spurious excitations in the NQS generated by the external magnetic field. This procedure allows us to select the state with $L_z=L$, and hence $J=J_z$, provided that $S$ and $L$ be aligned. 

Fig.~\ref{fig:B_field} illustrates the last two stages of the algorithm applied to $^7$Li, with a ground state of $J^\pi=3/2^+$. Panels (a) and (b) show that, towards the end of the second stage, the magnetic field generates components in the NQS with $J>3/2$, causing the expectation values of $J$ and $J_z$ to exceed the target values. However, upon switching off the magnetic field, the SR algorithm projects out the excited-states contamination, leading both $J$ and $J_z$ to converge to the desired values. Panel (c) shows that the external magnetic field enhances the expectation value of the magnetic moment, stabilizing it to a converged value when $B_z$ is set to zero.

The NQS wave function of $^{15}$O$(J^\pi=1/2^+)$ and $^{15}$N$(J^\pi=1/2^+)$  are a superposition of $J=1/2$ and $3/2$. To  extract the physical value of the magnetic moment for nuclei with $J<L+S$, we first project the NQS along all allowed $L_z$ for a given $L$. In order to converge to a non-maximal value of $L_z$, i.e. $|L_z|<L$, we add to the Hamiltonian the term $B_z|\hat{L}_z-\bar{L}_z|$, where $\bar{L}_z$ is the target $L_z$. Once the corresponding matrix elements $\langle L, L_z, S, S_z|\mu_z| L, L_z, S, S_z\rangle$ are computed, we take advantage of the Wigner-Eckart theorem to estimate the expectation value of the magnetic moment as defined in Eq.~(\ref{eq:mag_mom}) [see supplemental materials for more details].

\begin{figure}[htb]
    \centering
    \includegraphics[width=\columnwidth]{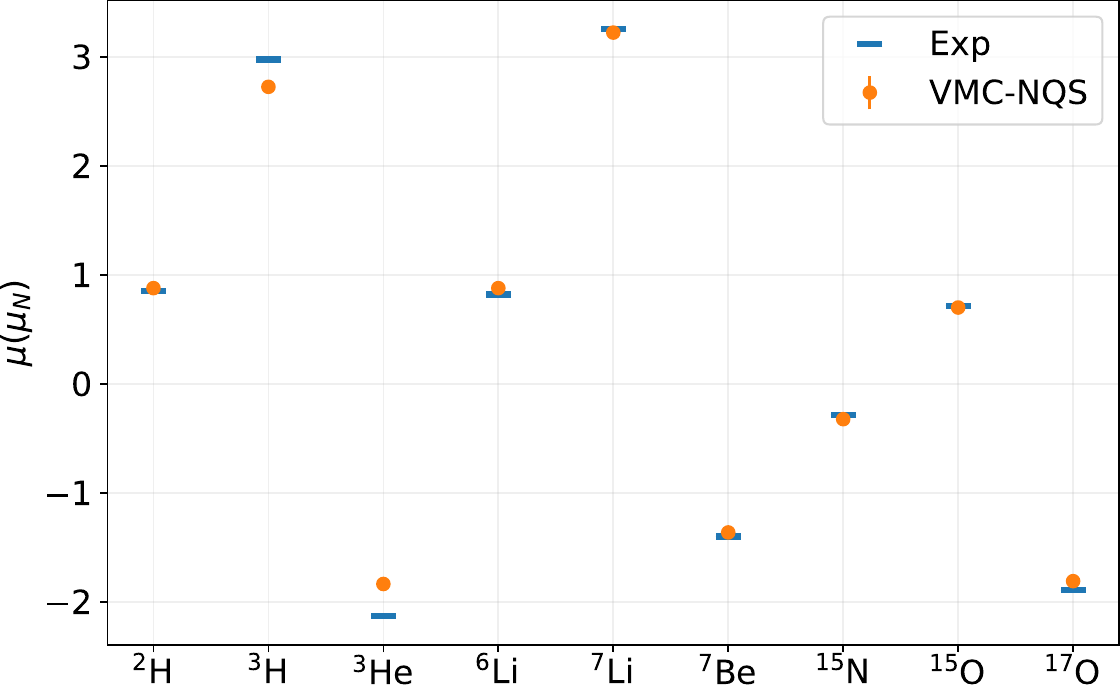}
    \caption{Magnetic moments of selected $A\leq 20$ nuclei obtained from VMC-NQS calculations compared with experimental data.}
    \label{fig:moments}
\end{figure}

As shown in Fig.~\ref{fig:moments}, theoretical calculations reproduce experimental data very well, implying that the input LO $\slashed{\pi}$EFT Hamiltonian encapsulates the essential elements of nuclear structure. This agreement also confirms the correctness of the algorithm discussed for selecting the polarization of the nucleus. Finally, it indicates that NSQ can learn the shell structure of the nucleus, which is genuinely self-emerging as it was not encoded in the variational ansatz. The minor discrepancies observed in $^3$H, and $^3$He, are consistent with GFMC, AFDMC and HH results~\cite{Pastore:2012rp,Gnech:2022,Martin:2023dhl} and may well be resolved when two-body currents are included in the calculation. 

\paragraph{Conclusions --} 
With the overarching aim of elucidating the essential aspects of nuclear binding, we accurately computed ground-state properties of atomic nuclei containing up to $A=20$ nucleons using as input the LO $\slashed{\pi}$EFT Hamiltonian ``o'' of Ref.~\cite{Schiavilla:2021dun}. To achieve precise solutions to the many-body Schrödinger equation, we employed a VMC method based on the versatile hidden-nucleon NQS framework, introduced in Ref.~\cite{Lovato:2022tjh}. Taking inspiration from condensed-matter studies~\cite{Pescia:2023mcc, Kim:2023fwy}, we significantly enhanced the convergence and expressive power of the original hidden-nucleon ansatz by employing equivariant backflow transformations based on MPNN to preprocess the single-nucleon coordinates and incorporate correlation effects.

In line with the findings of Ref.~\cite{Lu:2018bat}, which employed a spin-isospin independent, non-local, SU(4)-symmetric Hamiltonian, our computed binding energies typically exhibit deviations of only a few percent from experimental values. This level of precision is particularly remarkable considering the simplicity of the input Hamiltonian, especially when compared to the more sophisticated N$^2$LO chiral-EFT interactions that struggle to replicate ground-state energies with comparable accuracy~\cite{Martin:2023dhl}. Furthermore, the $\slashed{\pi}$EFT model ``o'' considered in this work --- albeit with a slightly different value for the three-body force regulator --- yields a neutron-matter equation of state that closely resembles the one obtained from the highly-realistic Argonne $v_{18}$ plus Urbana IX potentials~\cite{Fore:2022ljl}. This contrasts with the $\Delta$-full chiral-EFT Hamiltonian ``NV2+Ia,'' which provides an accurate energy spectrum for $A\leq 12$ nuclei~\cite{Piarulli:2017dwd}, yet falters in accurately modeling infinite neutron matter~\cite{Lovato:2022apd}. We ascribe the surprising accuracy of our ``essential'' Hamiltonian to a combination of two factors: i) the two-body potential reproduces peripheral, low-energy nucleon-nucleon scattering, and ii) the three-body force prevents nucleons from packing too closely, making low-momentum observables, such as binding energies and radii, insensitive to short-range details of the nuclear force.  

Unlike quantum many-body methods based on single-particle basis expansion~\cite{Barrett:2013nh,Caprio:2022mkg}, and akin to conventional QMC methods, NQS have no difficulties in capturing the slowly-decaying tails of nuclear wave functions, which play a crucial role in reproducing charge radii. Our theoretical calculations align reasonably well with experiments, although some discrepancies are present. Hence, similar to high-precision chiral-EFT forces~\cite{Soma:2019bso}, charge radii offer additional insights into nuclear dynamics not provided by binding energies alone~\cite{GarciaRuiz:2016ohj,Koszorus:2020mgn}.

Lastly, we introduced an innovative computational protocol to assess magnetic moments, amenable to NQS. It involves adding to the nuclear Hamiltonian an external magnetic field, which projects the NQS onto a specified value of the orbital angular momentum along the $z$ axis. Our computed magnetic dipole moments are in good agreement with experimental data; the discrepancies observed in $^3$H and $^3$He are likely to be resolved with the inclusion of two-body currents~\cite{Pastore:2012rp, Martin:2023dhl, Gnech:2022}. Beyond serving as an additional test for the underlying Hamiltonian, this concordance reveals how nucleons self-organize within shells, a structural aspect of the nucleus that is not explicitly embedded within the NQS ansatz but reveal its flexibility.

\paragraph{Acknowledgments -- } 
We are grateful to Giuseppe Carleo, Patrick Fasano, Jordan Fox, Antoine Georges, Alejandro Kievsky, Jane Kim, Tommaso Morresi,  Gabriel Pescia, Michele Viviani, and Robert B. Wiringa for many illuminating discussions. AL and BF are supported by the U.S. Department of Energy, Office of Science, Office of Nuclear Physics, under contracts DE-AC02-06CH11357, by the Department of Energy Early Career Award Program, by the NUCLEI SciDAC Program, and Argonne LDRD awards.
We gratefully acknowledge the computing resources provided on Swing, a high-performance computing cluster operated by the Laboratory Computing Resource Center at Argonne National Laboratory. 
We also acknowledge the use of resources of the National Energy Research Scientific Computing Center (NERSC), a U.S. Department of Energy Office of Science User Facility located at Lawrence Berkeley National Laboratory, operated under Contract No. DE-AC02-05CH11231 using NERSC award NP-ERCAP0023221.

\bibliography{biblio}

\appendix*
\section{Appendix: Wigner-Eckart}
In this Section, we outline the procedure used to compute the magnetic moments of nuclei with $J<L+S$, such as $^{15}$O and $^{15}$N. The starting point are the NQS with with good $L_z$, which are linear combinations of $J$, $J_z$ eigenstates
\begin{equation}
    |L,L_z;S,S_z\ket=\sum_{J,J_z}\bra L,L_z;S,S_z|J,J_z\ket |J ,J_z\ket\,,
\end{equation}
where $\bra L,L_z;S,S_z|J,J_z\ket$ are the Clebsch-Gordan coefficient~\cite{Workman:2022ynf}. We define the magnetic moments of these states as 
\begin{align}
    \mu_{L_z}=\bra L,L_z; S, S_z|\hat{\mu_z}|L;L_z; S, S_z\ket\,.\label{eq:mul}\,.
\end{align}
Exploiting the Wigner-Eckart theorem, these matrix elements can be expressed in terms of the reduced matrix elements $\bra J||\hat{\mu_z} ||J'\ket$ as 
\begin{align}    
\mu_{L_z}&=\sum_{J,J_z, J',J_z'}\bra L,L_z;S,S_z|J,J_z\ket \bra L,L_z;S,S_z|J',J'_z\ket\nonumber\\
&\times
    \frac{\bra J',J_z';1,0|J J_z\ket}{\sqrt{2J+1}}\bra J||\hat{\mu_z} ||J'\ket\,,\label{eq:linsys}\,,
\end{align}
with $\bra J||\hat{\mu_z} ||J'\ket=\bra J'||\hat{\mu_z} ||J\ket$. The latter can be found by solving a linear systems of equations, provided enough $\mu_{L_z}$ are independently evaluated. From the reduced matrix elements, it is immediate to obtain the physical ones by exploiting again the Wigner-Eckart theorem
\begin{align}
    \bra J,J_z|\hat{\mu}_z|J,J_z\ket=\frac{\bra J,J_z; 1,0|J,J_z\ket}{\sqrt{2J+1}}\bra J ||\hat{\mu}_z||J\ket\,.
\end{align}

The $^{15}$O and $^{15}$N nuclei are both $(J=1/2^+)$ and the NQS converges to $L=1$ and $S=1/2$. For each nucleus we calculated the matrix elements in Eq.~(\ref{eq:mul}) with $L_z=0,+1$, and $-1$ ($S_z$ is conveniently fixed in the Metropolis Hastings Sampling to the value $S_z = 1/2$). Solving the linear system generated by Eq.~(\ref{eq:linsys}), we obtain the magnetic moment matrix element for the physical state $J=1/2$ and $J_z=1/2$ as function of $\mu_{L_z}$, which has the compact expression 
\begin{align}
    \mu_{J=1/2}=\bra 1/2,1/2|\hat{\mu}_z|1/2,1/2\ket=\mu_0-\mu_{-1}-\frac{1}{3}\mu_{+1}\,.
\end{align}

\end{document}